\documentclass[journal]{IEEEtran}
\newtheorem{theorem}{Theorem}[section]

\newtheorem{lemma}[theorem]{Lemma}

\newtheorem{remark}{Remark}[section]

\newcommand{\R}{{\mathbb R}}

\newcommand{\C}{{\mathcal{C}}}
\newcommand{\scs}{{\mathcal{S}}}

\newcommand{\K}{{\mathcal{K}}}

\newcommand{\G}{{\mathcal{G}}}

\newcommand{\F}{{\cal F}}
\newcommand{\samp}{{\mathcal{X}}}

\newcommand{\ham}{{\Upsilon}}

\def\qt#1{\qquad\text{#1}}

\usepackage{verbatim}

\usepackage{cite}

\ifCLASSINFOpdf
\else
\fi
\usepackage[cmex10]{amsmath}
\usepackage{amssymb,amsfonts}
\begin{document}
%
\title{Covering Numbers for Convex Functions}
%
%
%

\author{Adityanand~Guntuboyina 
        and Bodhisattva~Sen
\thanks{A. Guntuboyina is with the Department
of Statistics, University of California, Berkeley,
CA 94720 USA e-mail: aditya@stat.berkeley.edu.}
\thanks{B. Sen is with the Department
of Statistics, Columbia University, New York,
NY 10027 USA e-mail: bodhi@stat.columbia.edu.}
}

%
%

\markboth{SUBMITTED TO IEEE TRANSACTIONS ON INFORMATION THEORY, 2012}%
{SUBMITTED TO IEEE TRANSACTIONS ON INFORMATION THEORY, 2012}
%



\maketitle

\begin{abstract}
In this paper we study the covering numbers of the space of convex and
uniformly bounded functions in multi-dimension. We find
optimal upper and lower bounds for the $\epsilon$-covering number of 
$\C([a, b]^d, B)$, in the $L_p$-metric, $1 \le p < \infty$, in terms
of the relevant constants, where $d \geq 1$, $a < b \in \mathbb{R}$,
$B>0$, and $\C([a,b]^d, B)$ denotes the set of all convex functions on
$[a, b]^d$ that are uniformly bounded by $B$. We summarize
previously known results on covering numbers for convex functions
and also provide alternate proofs of some known results. Our results
have direct implications in the study of rates of convergence of
empirical minimization procedures as well as optimal convergence
rates in the numerous convexity constrained function estimation problems.
\end{abstract}


%
\IEEEpeerreviewmaketitle

\section{Introduction}
%
%
%
%
\IEEEPARstart{E}{ver} since the work of~\cite{KolmogorovTihomirov},
covering numbers
(and their logarithms, known as metric entropy numbers) have been
studied extensively in a variety of disciplines. For a subset $\F$
of a metric space $(\samp, \rho)$, the $\epsilon$-{\it covering number}
$M(\F, \epsilon; \rho)$ is defined as the smallest
number of balls of radius $\epsilon$ whose union contains
$\F$. Covering numbers capture the \textit{size} of the underlying
metric space and play a central role in a number of areas in
information theory and statistics, including nonparametric function
estimation, density estimation, empirical processes and machine
learning.

In this paper we study the covering numbers of the space of convex and
uniformly bounded functions in multi-dimension. Specifically, we find
optimal upper and lower bounds for the $\epsilon$-covering number
$M(\C([a, b]^d, B), \epsilon; L_p)$, in the $L_p$-metric, $1 \le p <
\infty$, in terms of the relevant constants, where $d \geq 1$, $a, b
\in \mathbb{R}$, $B>0$, and $\C([a,b]^d, B)$ denotes the set of all
convex functions on $[a, b]^d$ that are uniformly bounded by $B$. We
also summarize previously known results on covering numbers for convex
functions. The special case of the problem when $d=1$ has been
recently established by Dryanov in~\cite[Theorem 3.1]{Dryanov}. Prior
to~\cite{Dryanov}, the only other result on the covering numbers of
convex functions is due to Bronshtein in \cite{Bronshtein76} (see also
\cite[Chapter 8]{Dudley99book}) who considered convex functions that
are uniformly bounded and uniformly Lipschitz with a {\it known}
Lipschitz constant under the $L_{\infty}$ metric. 

In recent years there has been an upsurge of interest in nonparametric
function estimation under convexity based constraints, especially in
multi-dimension. In general function estimation, it is well-known (see
e.g., \cite{Birge83, LeCam:73AnnStat, YangBarron, GuntuFdiv}) that the
covering numbers of the underlying function space can be used to
characterize optimal rates of convergence. They are also useful for
studying the rates of convergence of empirical minimization procedures 
(see e.g., \cite{VandegeerBook, BM93}). Our results have direct
implications in this regard in the context of understanding the
rates of convergence of the numerous convexity constrained function
estimators, e.g., the nonparametric least squares estimator of a
convex regression function studied in \cite{SS11, HD11}; the maximum
likelihood estimator of a log-concave density in multi-dimension
studied in \cite{SW10, CSS10, DSS11}. Also, similar problems that
crucially use convexity/concavity constraints to estimate sets have
also received recent attention in the statistical and machine learning
literature, see e.g., \cite{G11, GKM06}, and our results can be
applied in such settings.

The paper is organized as follows. In Section~\ref{motive}, we set up
notation and provide motivation for our main results, which are proved
in Section~\ref{main}. In Section~\ref{conclusion}, we draw some
connections to previous results on covering numbers for convex
functions and prove a related auxiliary result along with some
inequalities of possible independent interest.   

\section{Motivation}~\label{motive}
The first result on covering numbers for convex functions was proved
by Bronshtein in~\cite{Bronshtein76}, who considered convex functions
defined on a cube in $\R^d$ that are uniformly bounded and uniformly
Lipschitz. Specifically, let $\C([a, b]^d, B, \Gamma)$ denote the class of
real-valued convex functions defined on $[a, b]^d$ that are uniformly
bounded in absolute value by $B$ and uniformly Lipschitz with constant
$\Gamma$. In Theorem 6 of \cite{Bronshtein76}, Bronshtein proved that for 
$\epsilon$ sufficiently small, the logarithm of $M(\C([a,b]^d, B, \Gamma),
\epsilon; L_{\infty})$ can be bounded from above and below by a
positive constant (not depending on $\epsilon$) multiple of
$\epsilon^{-d/2}$. Note that the $L_\infty$ distance between two
functions $f$ and $g$ on $[a, b]^d$ is defined as $||f-g||_\infty :=
\sup_{x \in [a, b]^d} |f(x) - g(x)|$.

Bronshtein worked with the class $\C([a, b]^d, B, \Gamma)$ where
the functions are uniformly Lipschitz with constant
$\Gamma$. However, in convexity-based 
function estimation problems, one usually does not have a known
uniform Lipschitz bound on the unknown function class. This leads to
difficulties in the analysis of empirical minimization procedures via
Bronshtein's result. To the best of our knowledge, there does not
exist any other result on the covering numbers of convex functions
that deals with all $d \geq 1$ and does not require the Lipschitz
constraint.

In the absence of the uniformly Lipschitz constraint (i.e., if one
works with the class $\C([a, b]^d, B)$ instead of $\C([a, b]^d, B,
\Gamma)$), the covering numbers under the $L_{\infty}$ metric are
infinite. In other words, the space $\C([a, b]^d, B)$ is not totally
bounded under the $L_{\infty}$ metric. This can be seen, for example,
by noting that the functions 
\begin{equation*}
  f_j(t) := \max \left(0, 1-2^j t \right), \qt{for $t \in [0, 1]$},
\end{equation*}
are in $\C([0, 1], 1)$, for all $j \geq 1$, and satisfy
\begin{equation*}
 ||f_j - f_k||_{\infty} \geq |f_j(2^{-k}) - f_k(2^{-k})| = 1 - 2^{j-k}
 \geq 1/2,
\end{equation*}
for all $j < k$.

This motivated us to study the covering numbers of the class $\C([a,
b]^d, B)$ under a different metric, namely the $L_p$-metric for $1
\leq p < \infty$. We recall
that under the $L_p$-metric, $1 \le p < \infty$, the distance between
two functions $f$ and $g$ on $[a, b]^d$ is defined as
\begin{equation*}
||f-g||_p :=  \left( \int_{x \in [a, b]^d} |f(x) - g(x)|^p dx
\right)^{1/p}.
\end{equation*}
Our main result in this paper shows that if one works with the
$L_p$-metric as opposed to $L_{\infty}$, then the covering numbers of 
$\C([a, b]^d, B)$ are finite. Moreover, they are bounded from above
and below by constant multiples of $\epsilon^{-d/2}$ for sufficiently
small $\epsilon$.

\section{$L_p$--covering number bounds for $\C([a, b]^d,
  B)$}\label{main}
In this section, we prove upper and lower bounds for the 
$\epsilon$-covering number of $\C([a, b]^d, B)$ under the
$L_p$-metric, $1 \le p \le \infty$. Let us start by noting a simple
scaling identity that allows us to take $a = 0, b = 1$ and $B = 1$,
without loss of generality. For each $f \in \C([a, b]^d, B)$, let us
define $\tilde{f}$ on $[0, 1]^d$ by $\tilde{f}(x) := f(a \mathbf{1} +
(b-a)x)/B$, where $\mathbf{1} = (1, \dots, 1) \in
\mathbb{R}^d$. Clearly $\tilde{f} \in \C([0, 1]^d, 1)$ and, for $1
\leq p < \infty$, 
\begin{eqnarray*}
& &  B^p \int_{x \in [0, 1]^d} \left|\tilde{f}(x) - g(x) \right|^p dx  \\
 & = &  (b-a)^{-d}\int_{y \in [a, b]^d}  \left|f(y) - B g \left(\frac{y - a
       \mathbf{1}}{b-a} \right) \right|^p dy.
\end{eqnarray*}
for $g \in \C([0, 1]^d, 1)$. It follows that covering $f$ to within $\epsilon$ in the
$L_p$-metric on $[a, b]^d$ is equivalent to covering $\tilde{f}$ to
within $(b-a)^{-d/p}\epsilon/B$ in the $L_p$-metric on $[0,
1]^d$. Therefore, for $1 \leq p < \infty$,
\begin{equation}\label{uglyscale}
  M(\C([a, b]^d, B), \epsilon; L_p) = M(\C([0, 1]^d, 1), \epsilon'
; L_p),
\end{equation}
where $\epsilon' := (b-a)^{-d/p} \epsilon/B$.

\subsection{Upper Bound for $M(\C([a, b]^d, B), \epsilon; L_p)$}
\begin{theorem}\label{UpperBound}
Fix $1 \leq p < \infty$. There exist positive constants $c$ and
$\epsilon_0$, depending only on the dimension $d$ and $p$, such that,
for every $B > 0$ and $b > a$, we have 
  \begin{equation*}
    \log M \left(\C([a, b]^d, B), \epsilon; L_p \right) \leq c \left(
      \frac{\epsilon}{B(b-a)^{d/p}} \right)^{-d/2},
  \end{equation*}
for every $\epsilon \leq \epsilon_0 B (b-a)^{d/p}$.
\end{theorem}

The main ingredient in our proof of the above theorem is an extension
of Bronshtein's theorem to uniformly bounded convex functions having
different Lipschitz constraints in different directions. Specifically,
for $B \in (0, \infty)$, $\Gamma_i \in (0, \infty]$ and $a_i < b_i$ 
for $i = 1, \dots, d$, let $\C \left( \prod_{i=1}^d [a_i,
  b_i]; B; \Gamma_1, \dots, \Gamma_d \right)$ denote the set of all
real-valued convex functions $f$ on the rectangle $[a_1, b_1] \times
\dots \times [a_d, b_d]$ that are  uniformly bounded by $B$ and
satisfy:
\begin{eqnarray}\label{eq:CoordinateLip}
  && \left|f(x_1, \dots, x_{i-1}, x_i, x_{i+1}, \dots, x_d) \right. \nonumber \\
  &&  \left. - f(x_1,
    \dots, x_{i-1}, y_i, x_{i+1}, \dots, x_d) \right| \leq  \Gamma_i |x_i -  y_i|
\end{eqnarray}
for every $i = 1, \dots, d$; $x_i, y_i \in [a_i, b_i]$ and $x_j \in
[a_j, b_j]$ for $j \neq i$. In other words, the function $x \mapsto
f(x_1, \dots, x_{i-1}, x, x_{i+1}, \dots, x_d)$ is Lipschitz on $[a_i,
b_i]$ with constant $\Gamma_i$ for all $x_j \in [a_j, b_j], j \neq
i$. 

Clearly, the class $\C([a, b]^d, B, \Gamma)$ that Bronshtein studied
is contained in $\C([a, b]^d; B; \Gamma, \dots, \Gamma)$. Also, it is
easy to check that every function $f$ in $\C \left( 
  \prod_i [a_i,   b_i]; B; \Gamma_1, \dots, \Gamma_d \right)$ is
Lipschitz with respect to the Euclidean norm on $\prod_i [a_i, b_i]$
with Lipschitz constant $\sqrt{\Gamma_1^2 + \dots + \Gamma_d^2}$. 

Note that for $\Gamma_i = \infty$, the
inequality~\eqref{eq:CoordinateLip} is satisfied by every function
$f$. As a result, we have the equality $\C([a, b]^d, B) = \C([a,
b]^d;B; \infty, \dots, \infty)$. The following result gives an upper
bound for the $\epsilon$-covering number of $\C(\prod_i[a_i, b_i]; B;
\Gamma_1, \dots, \Gamma_d)$ and is the main ingredient in the proof of
Theorem~\ref{UpperBound}. Its proof is similar to Bronshtein's
proof~\cite[Proof of Theorem 6]{Bronshtein76} of his upper bound on
$\C([a, b]^d, B, \Gamma)$ and is included in Section~\ref{conclusion}. 

\begin{theorem}\label{difflip}
There exist positive constants $c$ and $\epsilon_0$, depending only on
the dimension $d$, such that for every positive $B, \Gamma_1, \dots,
\Gamma_d$ and rectangle $[a_1, b_1] \times \dots \times [a_d, b_d]$,
we have
  \begin{eqnarray}\label{difflip.eq}
    & & \log M \left(\C \left( \prod_{i=1}^d [a_i, b_i];B; \Gamma_1,
        \dots, \Gamma_d \right), \epsilon; L_{\infty} \right) \nonumber \\
       & & \;\;\;\;\;\;\;\;  \leq c
    \left(\frac{B + \sum_{i=1}^d \Gamma_i (b_i - a_i)}{\epsilon}
    \right)^{d/2},
  \end{eqnarray}
for all $0 < \epsilon \leq \epsilon_0 \{B + \sum_{i=1}^d \Gamma_i(b_i -
a_i)\}$.
\end{theorem}
\begin{remark}
  Note that the right hand side of~\eqref{difflip.eq} equals $\infty$
  unless $\Gamma_i < \infty$ for all $i = 1, \dots, d$. Thus,
  Theorem~\ref{difflip} is only meaningful when $\Gamma_i < \infty$
  for all $i = 1, \dots, d$. 
\end{remark}
\begin{remark}
  Because $\C([a, b]^d, B, \Gamma)$ is contained in $\C([a, b]^d; B;
  \Gamma_1, \dots, \Gamma_d)$, Theorem~\ref{difflip} includes
  Bronshtein's upper bound on $\C([a, b]^d, B, \Gamma)$ as a special
  case. Moreover, it gives explicit dependence of the upper bound on
  the constants $a, b, B$ and $\Gamma$. Bronshtein did not state the
  dependence on these constants. 
\end{remark}


We are now ready to prove Theorem~\ref{UpperBound} using
Theorem~\ref{difflip}. Here is the intuition behind the
proof. The class $\C([a, b]^d, B)$ can be thought of as an expansion
of the class $\C([a, b]^d; B; \Gamma_1, \dots, \Gamma_d)$ formed by the
removal of the $d$ Lipschitz constraints $\Gamma_1, \dots, \Gamma_d$
(or equivalently, by setting $\Gamma_1 = \dots = \Gamma_d =
\infty$). Instead of removing all these $d$ Lipschitz constraints at the
same time, we remove them sequentially one at a time. This is formally
accomplished by induction on the number of indices $i$ for which
$\Gamma_i = \infty$. Each step of the induction argument focuses on
the removal of one finite $\Gamma_i$ and is thus like solving the
one-dimensional problem. We consequently use Dryanov's ideas
from~\cite[Theorem  3.1]{Dryanov} to solve this quasi one-dimensional
problem which allows us to complete the induction step.   

\begin{IEEEproof}[Proof of Theorem~\ref{UpperBound}]
The scaling identity (\ref{uglyscale})
lets us take $a = 0, b = 1$ and $B = 1$.

We shall prove that there exist positive constants $c$ and
$\epsilon_0$, depending only on $d$ and $p$, such that for
every $\Gamma_i \in (0, \infty]$, we have
\begin{eqnarray}\label{indu}
    & & \log M \left(\C \left([0, 1]^d;1; \Gamma_1, \dots, \Gamma_d
      \right) ;\epsilon; L_p \right) \nonumber \\
      && \;\;\;\;\; \leq  c \left(\frac{2 +
        \sum_{i=1}^d \Gamma_i \left\{\Gamma_i < \infty
        \right\}}{\epsilon} \right)^{d/2},
\end{eqnarray}
for $0 < \epsilon \leq \epsilon_0$. Note that this proves the theorem
because we can set $\Gamma_i = \infty$ for all $i = 1, \dots, d$. Our
proof will involve induction on $l$: the number of indices $i$
for which $\Gamma_i = \infty$.

For $l = 0$, i.e., when $\Gamma_i < \infty$ for all $i = 1, \dots,
d$,~\eqref{indu} is a direct consequence of Theorem~\ref{difflip}. In
fact, in this case,~\eqref{indu} also holds for $p = \infty$. Suppose
now that~\eqref{indu} holds for all $l < k$ for some $k \in \{1,
\dots, d\}$. We shall then verify it for $l = k$. Fix $\Gamma_i
\in (0, \infty]$ such that exactly $k$ of them equal infinity. Without
loss of generality, we assume that $\Gamma_1 = \dots = \Gamma_{k} =
\infty$ and $\Gamma_i < \infty$ for $i > k$. For every sufficiently
small $\epsilon > 0$, we shall exhibit an $\epsilon$-cover of $\C([0,
1]^d; 1; \infty, \dots, \infty, \Gamma_{k+1}, \dots, \Gamma_d)$ in the
$L_p$-metric whose cardinality has logarithm bounded from above by a
constant multiple of $(\sum_{i > k} \Gamma_i + 2)^{d/2}
\epsilon^{-d/2}$. Note that for $k = d$, the term $\sum_{i > k}
\Gamma_i$ equals zero. For convenience, let us denote the class $\C([0,
1]^d; 1; \infty, \dots, \infty, \Gamma_{k+1}, \dots, \Gamma_d)$ by
$\G$ in the rest of this proof. 

Let
\begin{equation}\label{condc}
  u := \exp \left(-2(p+1)^2(p+2) \log 2 \right) ~~ \text{ and } ~~ v :=
  1-u.
\end{equation}
Fix $\eta > 0$ and choose an integer $A$ and $\delta_1, \dots,
\delta_{A+1}$ such that
\begin{equation*}
  \eta^p = \delta_1 < \dots < \delta_{A} < u \leq \delta_{A+1}.
\end{equation*}
For every two functions $f$ and $g$ on $[0, 1]^d$, we can obviously
decompose the integral $\int |f-g|^p$ as
\begin{eqnarray*}
&& \int_{[0, 1]^d} |f-g|^p = \int_{[0, u] \times [0, 1]^{d-1}}
  |f-g|^p  \\
  & & \;\;\;\;\; + \int_{[u, v] \times [0, 1]^{d-1}}
  |f-g|^p  + \int_{[v, 1] \times [0, 1]^{d-1}} |f-g|^p.
\end{eqnarray*}
Also,
\begin{eqnarray*}
 && \int_{[0,u] \times [0,1]^{d-1}} |f-g|^p \leq
  \int_{[0,\delta_1] \times [0, 1]^{d-1}} |f-g|^p \\
  & & \;\;\;\;\; +  \sum_{m=1}^A \int_{[\delta_m, \delta_{m+1}]
    \times [0, 1]^{d-1}} |f-g|^p.
\end{eqnarray*}
For a fixed $m = 1, \dots, A$, consider the problem of covering the
functions in $\G$ on the rectangular strip $[\delta_m, \delta_{m+1}]
\times [0,1]^{d-1}$. Clearly,
\begin{equation}\label{eq:keytemp}
  \int_{[\delta_m, \delta_{m+1}] \times [0,1]^{d-1}} |f-g|^p  =
  (\delta_{m+1} - \delta_m)  \int_{[0,
    1]^d} |\tilde{f}-\tilde{g}|^p
\end{equation}
where, for $x = (x_1, \dots, x_d) \in [0, 1]^d$,
\begin{eqnarray*}
  \tilde{f}(x) := f(\delta_m + (\delta_{m+1} - \delta_m)x_1, x_2,
  \dots, x_d), \\
~~  \text{ and } ~~   \tilde{g}(x) := g(\delta_m + (\delta_{m+1} -
\delta_m)x_1, x_2, \dots, x_d).
\end{eqnarray*}
By convexity, the restriction of every function $f$ in $\G$ 
to $[\delta_m, \delta_{m+1}] \times [0, 1]^{d-1}$  belongs to the class:
\begin{equation*}
\C([\delta_m, \delta_{m+1}] \times [0, 1]^{d-1}; 1; 2/\delta_m,
\infty, \dots, \infty, \Gamma_{k+1},\ldots,\Gamma_d)
\end{equation*}
Consequently, the corresponding function $\tilde f$ belongs to
\begin{equation*}
\C([0,1]^{d}; 1; 2(\delta_{m+1} - \delta_m)/\delta_m,
\infty, \dots, \infty, \Gamma_{k+1},\ldots, \Gamma_d). 
\end{equation*}
Because $2(\delta_{m+1} - \delta_m)/\delta_m < \infty$, we can use the
induction hypothesis to assert the existence of positive constants
$\epsilon_0$ and $c$, depending only on $d$ and $p$, such that for
every positive real number $\alpha_m \leq \epsilon_0$, there exists an 
$\alpha_m$-cover of $\C([0, 1]^{d}; 1; 2(\delta_{m+1} -
\delta_m)/\delta_m, \infty, \dots, \infty, \Gamma_{k+1},\ldots,
\Gamma_d))$ in the $L_p$-metric on $[0,1]^{d}$ of size smaller than
\begin{eqnarray*}
  & & \exp \left(c \alpha_m^{-d/2} \left(2 + \frac{2(\delta_{m+1}-
        \delta_m)}{\delta_m} + \sum_{i>k} \Gamma_i \right)^{d/2}
  \right) \\
  & & \;\;\;\;\;\leq \exp \left(c \left(2 + \sum_{i > k} \Gamma_i \right)^{d/2}
  \left(\frac{\delta_{m+1}}{\delta_m \alpha_m} \right)^{d/2} \right).
\end{eqnarray*}
By covering the functions in $\G$ by the constant function 0 on $[0,
\delta_1] \times [0, 1]^{d-1}$ and up to 
$\alpha_m$ in the $L_{p}$-metric on $[\delta_m, \delta_{m+1}]
\times [0, 1]^{d-1}$ for $m = 1, \dots, A$, we obtain a cover of the
restriction of the functions in $\G$ to the set $[0, u] \times
[0,1]^{d-1}$ in $L_p$-metric having coverage $S_1^{1/p}$ and
cardinality bounded from above by $\exp (S_2)$ where
\begin{eqnarray}\label{cov}
& & S_1 := \delta_1 + \sum_{m=1}^A \alpha_m^p (\delta_{m+1} - \delta_m) ~ \text{ and }\nonumber \\
&&  S_2 := c \left( \sum_{i > k} \Gamma_i  + 2
\right)^{d/2} \sum_{m=1}^A \left(\frac{\delta_{m+1}}{\delta_m
    \alpha_m} \right)^{d/2}.
\end{eqnarray}
Suppose now that
\begin{eqnarray*}
  \delta_m & := & \exp \left(p\left(\frac{p+1}{p+2} \right)^{m-1} \log
    \eta \right) ~~ \text{ and } \nonumber \\
     ~~ \alpha_m & := & \eta \exp \left( - p
      \frac{(p+1)^{m-2}}{(p+2)^{m-1}}  \log \eta  \right),
\end{eqnarray*}
for $m = 1, \dots, A+1$, where $A$ is the largest integer such that
\begin{equation*}
  \exp \left(p\left(\frac{p+1}{p+2} \right)^{A-1} \log
    \eta \right) < u.
\end{equation*}
Then,
\begin{eqnarray*}
  S_1 & = & \delta_1 + \sum_{m=1}^A \alpha_m^p \left( \delta_{m+1} -
  \delta_m \right) \\
  & \leq & \delta_1 + \sum_{m=1}^A \alpha_m^p
  \delta_{m+1} = \eta^p \left( 1 + \sum_{m=1}^A \zeta_m^2 \right),
\end{eqnarray*}
and
\begin{equation*}
  S_2 = c \left(\frac{ \sum_{i > k} \Gamma_i  + 2}{\eta}
  \right)^{d/2}  \sum_{m=1}^A \zeta^d_m,
\end{equation*}
where
\begin{equation*}
\zeta_m := \sqrt{\frac{\eta \delta_{m+1}}{\delta_m \alpha_m}} = \exp
  \left(\frac{p}{2(p+1)^2} \frac{(p+1)^m}{(p+2)^m} \log \eta
  \right).
\end{equation*}

Note that if $\eta \leq 1$, then $\log \eta \leq 0$ which
implies $\zeta_m \leq 1$. Also, for $m = 2, \dots, A$, we have
\begin{eqnarray*}
  \frac{\zeta_m}{\zeta_{m-1}} & = & \exp \left(\frac{-p \log
      \eta}{2(p+1)^2(p+2)} \left( \frac{p+1}{p+2} \right)^{m-1}
  \right) \\
  &\geq & \exp \left(\frac{-p \log
      \eta}{2(p+1)^2(p+2)} \left( \frac{p+1}{p+2} \right)^{A-1}
  \right) \\
&= &\exp \left(\frac{-\log \delta_A}{2(p+1)^2(p+2)} \right) \\
&> & \exp \left(\frac{-\log u}{2(p+1)^2(p+2)} \right) = 2,
\end{eqnarray*}
where we have used $\delta_A < u$ and the fact that $u$ has the
expression~\eqref{condc}. Therefore $\zeta_m \geq 2 \zeta_{m-1}$ which
can be rewritten as
\begin{equation*}
\zeta_m^r \leq \frac{2^r}{2^r-1}\left(\zeta_m^r - \zeta_{m-1}^r
\right) \qt{for every $r \geq 1$}.
\end{equation*}
Thus,
\begin{eqnarray*}
  \sum_{m=1}^A \zeta_m^r & \leq & \zeta_1^r + \frac{2^r}{2^r - 1}
  \sum_{m=2}^A \left(\zeta_m^r - \zeta_{m-1}^r \right) \\
  & = & \frac{1}{2^r - 1} \left(2^r \zeta_A^r - \zeta_1^r \right) \leq
  \frac{2^r}{2^r-1}.
\end{eqnarray*}
Using this for $r = 2$ and $r = d$, we deduce that
\begin{equation*}
  S_1 \leq \frac{7}{3} \eta^p ~~ \text{ and } ~~ S_2 \leq
  \frac{2^dc}{2^d - 1} \left(\frac{\sum_{i > k} \Gamma_i +
      2}{\eta} \right)^{d/2}.
\end{equation*}
An exactly similar analysis can be done now to cover the restrictions
of the functions in $\G$ to the set $[v, 1] \times [0, 1]^{d-1}$
having the same coverage 
$S_1^{1/p}$ and same cardinality bounded by $\exp(S_2)$. For $[u, v]
\times [0, 1]^{d-1}$, we note, by convexity, that the restrictions of
functions in $\G$ to the set
$[u, v] \times [0, 1]^{d-1}$ belong to $\C([u, v] \times [0, 1]^{d-1};
1; 2/u, \infty, \dots, \infty, \Gamma_{k+1}, \dots, \Gamma_d)$. By the
induction hypothesis, 
there exist constants $c$ and $\epsilon_0$, depending only on $d$ and
$p$, such that for all $\eta \leq \epsilon_0$, one can get a
$\epsilon$-cover of $\C([u, v] \times [0, 1]^{d-1}; 1; 2/u, \infty,
\dots, \infty, \Gamma_{k+1}, \dots, \Gamma_d)$ in the $L_p$-metric
having cardinality smaller than 
\begin{eqnarray*}
  \exp \left(c \eta^{-d/2} \left(2 + \frac{2}{u} + \sum_{i>k}
  \Gamma_i \right)^{d/2}  \right) \\
  \leq \exp \left(c \left(\frac{2}{u} \right)^{d/2}
  \left(\frac{\sum_{i>k} \Gamma_i + 2}{\eta} \right)^{d/2} \right).
\end{eqnarray*}
Observe that $u$ only depends on $p$. By combining the covers of the
restrictions of functions in $\G$ to these three strips $[0, u] \times
[0, 1]^{d-1}$, $[u, v] 
\times [0, 1]^{d-1}$ and $[v, 1] \times [0, 1]^{d-1}$, we obtain, for
$\eta \leq \epsilon_0$, a cover of $\G$ in the $L_p$-metric having
coverage at most 
\begin{equation*}
  \left(\frac{7}{3} \eta^p + \frac{7}{3} \eta^p + \eta^p
  \right)^{1/p} = \left(\frac{17}{3} \right)^{1/p} \eta
\end{equation*}
and cardinality at most
\begin{equation*}
  \exp \left(c \left(\frac{2^{d+1}}{2^d - 1} + \frac{2^{d/2}}{u^{d/2}}
    \right) \left(\frac{\sum_{i > k} \Gamma_i + 2}{\eta}
    \right)^{d/2}  \right).
\end{equation*}
By relabelling $(17/3)^{1/p} \eta$ as $\epsilon$, we have proved that
for $\epsilon \leq (3/17)^{1/p} \epsilon_0$,
\begin{eqnarray*}
  && \log M(\G; \epsilon; L_p) \\
  && \leq c \left(\frac{17}{3} \right)^{d/(2p)} \left(\frac{2^{d+1}}{2^d
      - 1} + \frac{2^{d/2}}{u^{d/2}} \right) \left(\frac{\sum_{i > k}
      \Gamma_i + 2}{\epsilon} \right)^{d/2}.
\end{eqnarray*}
This proves~\eqref{indu} for all $\Gamma_1, \dots, \Gamma_d$ such that
exactly $k$ of them equal $\infty$. The proof is complete by
induction.
\end{IEEEproof}
\begin{remark}
  The argument used in the induction step above involved splitting the
  interval $[0, 1]$ into the three intervals $[0, u], [u, v]$ and $[v,
  1]$, and then subsequently splitting the interval $[0, u]$ into 
  smaller subintervals. We have borrowed this idea from
  Dryanov~\cite[Proof of Theorem 3.1]{Dryanov}. We must mention
  however that Dryanov uses a  more elaborate argument to bound sums
  of the form $S_1$ and $S_2$. Our way of controlling $S_1$ and
  $S_2$ is much simpler which shortens the argument considerably. 
\end{remark}

\subsection{Lower bound for $M(\C([a, b]^d, B), \epsilon; L_p)$}
\begin{theorem}\label{thm:Lowerbound}
There exist positive constants $c$ and $\epsilon_0$, depending only on
the dimension $d$, such that for every $p \geq 1$, $B > 0$ and $b >
a$, we have 
  \begin{equation*}
    \log M \left(\C ([a, b]^d, B), \epsilon; L_p \right)
    \geq c \left(\frac{\epsilon}{B(b-a)^{d/p}} \right)^{-d/2},
  \end{equation*}
for $\epsilon \leq \epsilon_0 B (b-a)^{d/p}$.
\end{theorem}


\begin{IEEEproof}
As before, by the scaling identity~\eqref{uglyscale}, we take $a = 0$, 
$b = 1$ and $B = 1$. For functions defined on $[0, 1]^d$, the
$L_p$-metric, $p > 1$, is larger than $L_1$. We will thus take $p = 1$
in the rest of this proof. We prove that for $\epsilon$ sufficiently
small, there exists an $\epsilon$-packing subset of $\C([0, 1]^d, 1)$,
under the $L_1$-metric, of cardinality larger than a constant multiple
of $\epsilon^{-d/2}$. By a packing subset of $\C([0, 1]^d, 1)$, we mean a
subset $F$ satisfying $||f - g||_1 \geq \epsilon$ whenever $f, g \in
F$ with $f \neq g$. 

Fix $0 < \eta \leq 4 (2+\sqrt{d-1})^{-2}$ and let $k := k(\eta)$ be
the positive integer satisfying
\begin{equation}\label{kdef}
  k \leq \frac{2\eta^{-1/2}}{2+\sqrt{d-1}} < k+1 \leq 2k.
\end{equation}

Consider the intervals $I(i) = [u(i), v(i)]$ for $i = 1, \dots, k$, such that
\begin{enumerate}
\item $0 \leq u(1) < v(1) \leq u(2) < v(2) \leq \dots \leq u(k) < v(k) \leq 1$,
\item $v(i) - u(i) = \sqrt{\eta}$, for $i = 1, \dots, k$,
\item $u(i+1) - v(i) = \frac{1}{2}\sqrt{\eta(d-1)}$ for $i = 1, \dots,
  k-1$.
\end{enumerate}

Let $\scs$ denote the set of all $d$-dimensional cubes of the form
$I(i_1) \times \dots \times I(i_d)$ where $i_1, \dots, i_d \in \{1,
\dots, k\}$. The cardinality of $\scs$, denoted by $|\scs|$, is
clearly $k^d$.

For each $S \in \scs$ with $S = I(i_1) \times \dots \times I(i_d)$
where $I(i_j) = [u(i_j), v(i_j)]$, let us define the function $h_S: [0,1]^d \rightarrow \mathbb{R}$ as
\begin{eqnarray}\label{eq:hS-f0}
  h_S(x) & = & h_S(x_1,\ldots,x_d) \nonumber \\
  & := & \frac{1}{d} \sum_{j=1}^d \left[u^2(i_j) + \{v(i_j) +
    u(i_j)\} \{x_j - u(i_j) \} \right] \nonumber \\
    & = & f_0(x) + \frac{1}{d} \sum_{j=1}^d  \{x_j-u(i_j)\} \{v(i_j) - x_j\},
\end{eqnarray}
where $f_0(x) := \frac{1}{d} \left(x_1^2 + \dots + x_d^2 \right)$, for $x \in [0, 1]^d$.
The functions $h_S, S \in \scs$ have the following four key
properties:
\begin{enumerate}
\item $h_S$ is affine and hence convex.
\item For every $x \in [0, 1]^d$, we have $h_S(x) \leq h_S(1, \dots,
  1) \leq 1$.
\item For every $x \in S$, we have $h_S(x) \geq f_0(x)$. This is
  because whenever $x \in S$, we have $u(i_j) \leq x_j \leq v(i_j)$
  for each $j$, which implies $\{x_j - u(i_j)\}\{v(i_j) - x_j\} \geq 0$.
\item Let $S, S' \in \scs$ with $S \neq S'$. For every $x \in S'$, we
  have $h_S(x) \leq f_0(x)$. To see this, let $S' = I(i'_1) \times
  \dots \times I(i'_d)$ with $I(i'_j) = [u(i'_j), v(i'_j)]$. Let $x
  \in S'$ and fix $1 \leq j \leq d$. If $I(i_j) = I(i_j')$, then $x_j
  \in I(i_j) = [u(i_j), v(i_j)]$ and hence
  \begin{eqnarray*}
    \{x_j - u(i_j)\}\{v(i_j) - x_j\} \leq \frac{\{v(i_j) - u(i_j)\}^2}{4} =
    \frac{\eta}{4}.
  \end{eqnarray*}
  If $I(i_j) \neq I(i_j')$ and $u(i'_j) < v(i'_j) < u(i_j) < v(i_j)$,
  then
  \begin{eqnarray*}
    && \{x_j - u(i_j)\}\{v(i_j) - x_j\} \\
    & & \;\;\;\;\;\;\; \leq -\{u(i_j) - v(i'_j)\}^2 =
    -\frac{d-1}{4} \eta.
  \end{eqnarray*}
The same above bound holds if $u(i_j) < v(i_j) < u(i'_j) < v(i'_j)$.
Because $S \neq S'$, at least one of $i_j$ and $i'_j$ will be
different. Consequently,
\begin{align*}
  h_S(x) &= f_0(x) + \sum_{j} \{x_j - u(i_j)\}\{v(i_j) - x_j\}  \\
&\leq f_0(x) + \sum_{j: i_j = i'_j} \frac{\eta}{4} - \sum_{j : i_j
  \neq i'_j} (d-1) \frac{\eta}{4} \leq f_0(x).
\end{align*}
\end{enumerate}

Let $\{0, 1\}^{\scs}$ denote the collection of all $\{0, 1\}$-valued
functions on $\scs$. The cardinality of $\{0, 1\}^{\scs}$ clearly
equals $2^{|\scs|}$ (recall that $|\scs| = k^d$).

For each $\theta \in \{0, 1\}^{\scs}$, let
\begin{equation*}
  g_{\theta}(x) := \max \left( \max_{S \in \scs: \theta(S) = 1}
    h_S(x), f_0(x) \right).
\end{equation*}
The first two properties of $h_S, S \in \scs$ ensure that $g_{\theta}
\in \C([0, 1]^d, 1)$. The last two properties imply that
\begin{equation*}
  g_{\theta}(x) = h_S(x) \theta(S) + f_0(x) (1 - \theta(S)) \qt{for $x
    \in S$}.
\end{equation*}
We now bound from below the $L_1$ distance between $g_{\theta}$ and
$g_{\theta'}$ for $\theta, \theta \in \{0, 1\}^{\scs}$. Because the interiors of the
cubes in $\scs$ are all disjoint, we can write
\begin{eqnarray*}
  && ||g_{\theta} - g_{\theta'}||_1 \geq \sum_{S \in \scs} \int_{x \in S}
  \left|g_{\theta}(x) - g_{\theta'}(x)  \right| dx \\
  && = \sum_{S \in \scs}
  \left\{\theta(S) \neq \theta'(S) \right\} \int_{x \in S} |h_S(x) -
  f_0(x)| dx.
\end{eqnarray*}
Note that from~(\ref{eq:hS-f0}) and by symmetry, the value of integral
\begin{equation*}
  \zeta := \int_{x \in S} |h_S(x) - f_0(x)| dx
\end{equation*}
is the same for all $S \in \scs$. We have thus shown that
\begin{equation}\label{nopain}
  ||g_{\theta} - g_{\theta'}||_1 \geq \zeta \ham(\theta, \theta')
  \qt{for all $\theta, \theta' \in \{0, 1\}^{\scs}$},
\end{equation}
where $\ham(\theta, \theta') := \sum_{S \in \scs} \left\{\theta(S)
  \neq \theta'(S) \right\}$ denotes the Hamming distance.

The quantity $\zeta$ can be computed in the following way. Let $S =
I(i_1) \times \dots \times I(i_d)$ where $I(i_j) = [u(i_j),
v(i_j)]$. We write
\begin{equation*}
  \zeta = \int_{u(i_1)}^{v(i_1)} \dots \int_{u(i_d)}^{v(i_d)}
  \frac{1}{d} \sum_{j=1}^d \{x_j - u(i_j)\}\{v(i_j) - x_j\} dx_d \dots
  dx_1.
\end{equation*}
By the change of variable $y_j = \{x_j - u(i_j)\}/\{v(i_j) - u(i_j)\}$ for
$j = 1, \dots, d$, we get
\begin{equation*}
  \zeta = \prod_{j=1}^d \{v(i_j) - u(i_j)\} \int_{[0,1]^d} \frac{1}{d}
  \sum_{j=1}^d \{v(i_j) - u(i_j)\}^2 y_j(1-y_j) dy.
\end{equation*}
Recalling that $v(i) - u(i) = \sqrt{\eta}$ for all $i = 1, \dots, k$,
we get $\zeta = \eta^{d/2} \eta \gamma_d$ where
\begin{equation*}
  \gamma_d := \int_{[0,1]^d} \frac{1}{d} \sum_{j=1}^d y_j(1-y_j)dy.
\end{equation*}
Note that $\gamma_d$ is a constant that depends on the dimension $d$
alone. Thus, from~\eqref{nopain}, we deduce
\begin{equation}
  \label{pace}
  ||g_{\theta} - g_{\theta'}||_1 \geq \gamma_d \eta^{d/2} \eta
  \ham(\theta, \theta')
\end{equation}
for all $\theta, \theta' \in \{0, 1\}^{\scs}$. We now use the Varshamov-Gilbert lemma (see e.g., \cite[Lemma
4.7]{Massart03Flour}) which asserts the existence of a subset $W$ of
$\{0, 1\}^{\scs}$ with cardinality, $|W| \geq \exp(|\scs|/8)$ such that
$\ham(\tau, \tau') \geq |\scs|/4$ for all $\tau, \tau' \in W$ with $\tau
\neq \tau'$. Thus, from~\eqref{pace} and~\eqref{kdef}, we get that for
every $\tau, \tau' \in W$ with $\tau \neq \tau'$,
\begin{equation*}
  ||g_{\theta} - g_{\theta'}||_1 \geq \gamma_d \eta^{d/2} \eta
  \frac{|\scs|}{4} = \frac{\gamma_d}{4} \eta^{d/2} \eta k^d \geq c_1
  \eta
\end{equation*}
where $c_1 := \frac{\gamma_d}{4}(2+\sqrt{d-1})^{-d}$. Taking $\epsilon := c_1 \eta$, we have obtained for $\epsilon
\leq \epsilon_0 := 4c_1(2+\sqrt{d-1})^{-2}$, an $\epsilon$-packing
subset of $\C([0,1]^d, 1)$ of size $M := |W|$ where
\begin{eqnarray*}
  \log M & \geq & \frac{|\scs|}{8} = \frac{k^d}{8} \geq
  \frac{(2+\sqrt{d-1})^{-d}}{8} \eta^{-d/2} \\ 
  & =& \frac{c_1^{d/2}}{8(2+\sqrt{d-1})^d} \epsilon^{-d/2} =
  c\epsilon^{-d/2},
\end{eqnarray*}
where $c$ depends only on the dimension $d$. This completes the proof.
\end{IEEEproof}
\begin{remark}
The explicit packing subset constructed in the above proof consists of functions that can be viewed as perturbations of the quadratic function $f_0$. Previous lower bounds on the covering numbers of convex functions in~\cite[Proof of Theorem 6]{Bronshtein76} and~\cite[Section 2]{Dryanov} (for $d =1$) are based on
perturbations of a function whose graph is a subset of a sphere; a more complicated convex function than $f_0$.  The perturbations of $f_0$ in the above proof can also be used to simplify the lower bound arguments in those papers.
\end{remark}

\section{Distances between convex functions, and their
  epigraphs}\label{conclusion} 
One of the aims of this section is to provide the proof of
Theorem~\ref{difflip}. Our strategy for the proof of
Theorem~\ref{difflip} is similar to Bronshtein's proof of the
upper bound on $M(\C([a,b]^d, B, \Gamma), \epsilon; L_{\infty})$. The
proof involves the following ingredients:
\begin{enumerate}
\item An inequality between the $L_{\infty}$ distance between two
  convex functions and the Hausdorff distance between their
  epigraphs. 
\item The result of Bronshtein~\cite{Bronshtein76} for the covering
  numbers of convex sets in the Hausdorff metric.   
\end{enumerate}

For a convex function $f$ on $[0, 1]^d$ and $B > 0$, let us define the
epigraph $V_f(B)$ of $f$ by 
\begin{eqnarray*}
  V_f(B) := \left\{(x_1, \dots, x_d, x_{d+1}): (x_1, \dots, x_d) \in [0,
    1]^d \right. \\
   \left. \text{     and    } f(x_1, \dots, x_d) \leq x_{d+1} \leq B
  \right\}. 
\end{eqnarray*}
If $f \in \C([0, 1]^d, B)$, then clearly
\begin{equation*}
  x_1^2 + \dots + x_d^2 + x_{d+1}^2 \leq 1 + \dots + 1 + B^2 = d + B^2
\end{equation*}
for every $(x_1, \dots, x_{d+1}) \in V_f(B)$. Therefore, for every $f
\in \C([0, 1]^d, B)$, its epigraph $V_f(B)$ is contained in the
$(d+1)$-dimensional ball of radius $\sqrt{d+B^2}$ centered at the
origin. The following inequality relates the $L_{\infty}$ distance
between two functions in $\C([0, 1]^d; B; \Gamma_1, \dots, \Gamma_d)$
to the Hausdorff distance between their epigraphs. The Hausdorff
distance between two compact, convex sets $C$ and $D$ in Euclidean
space is defined by 
\begin{equation*}
  \ell_H(C, D) := \max \left(\sup_{x \in C} \inf_{y \in D} |x-y|,
    \sup_{x \in D} \inf_{y \in C} |x-y|  \right),
\end{equation*}
where $|\cdot|$ denotes Euclidean distance. 
\begin{lemma}\label{infvf}
For every pair of functions $f$ and $g$ in $\C([0, 1]^d; B;
\Gamma_1, \dots, \Gamma_d)$, we have 
  \begin{equation*}
    ||f-g||_{\infty} \leq \ell_H(V_f(B), V_g(B)) \sqrt{1 + \Gamma_1^2
      + \dots + \Gamma_d^2}. 
  \end{equation*}
\end{lemma}
\begin{IEEEproof}
We can clearly assume that $\Gamma_i < \infty$ for all $i = 1, \dots,
d$. Fix $f, g \in \C([0,1]^d; B; \Gamma_1, \dots, \Gamma_d)$
and let $\ell_H(V_f(B), V_g(B)) = \rho$. Fix $x \in [0,1]^d$ with
$f(x) \neq g(x)$. Suppose, without loss of
generality, that $f(x) < g(x)$. Now $(x, f(x)) \in V_f(B)$ and because
$\ell_H(V_f(B), V_g(B)) = \rho$, there exists $(x', y') \in V_g(B)$
with $|(x, f(x)) - (x', y')| \leq \rho$. Because $f(x) < g(x)$, the
point $(x, f(x))$ lies outside $V_g(B)$ and using the convexity of
$V_g(B)$ we can take $y' = g(x')$. Therefore,
\begin{align*}
\;\;\; 0 & \leq  g(x) - f(x) \\ 
&= g(x) - g(x') + g(x') - f(x) \\
&\leq |x-x'| \sqrt{\Gamma^2_1 + \dots + \Gamma^2_d} + |g(x') - f(x)| \\
&\leq \sqrt{\Gamma_1^2 + \dots + \Gamma_d^2 + 1} \sqrt{|x-x'|^2 +
  |g(x') - f(x)|^2} \\
&= \sqrt{\Gamma_1^2 + \dots + \Gamma_d^2 + 1}~ |(x, f(x)) - (x', y')| \\
& \leq \rho \sqrt{\Gamma_1^2 + \dots + \Gamma_d^2 + 1},
\end{align*}
where the second last inequality follows from the Cauchy-Scwarz (C-S)
inequality. Lemma~\ref{infvf} now follows because $x \in
[0,1]^d$ is arbitrary in the above argument.  
\end{IEEEproof}

The proof of Theorem~\ref{difflip}, given below, is based on
Lemma~\ref{infvf} and the following result on covering numbers of
convex sets proved in~\cite{Bronshtein76}. For $\Gamma > 0$, let
$\K^{d+1}(\Gamma)$ denote the set of all compact, convex subsets of
the ball in $\R^{d+1}$ of radius $\Gamma$ centered at the origin. In
Theorem 3 (and Remark 1) of \cite{Bronshtein76}, Bronshtein proved
that there exist positive constants $c$ and $\epsilon_0$, depending
only on $d$, such that 
\begin{equation}\label{goodbronsh}
 \log M(\K^{d+1}(\Gamma), \epsilon; \ell_H) \leq c
 \left(\frac{\Gamma}{\epsilon} \right)^{d/2} \qt{for $\epsilon \leq
   \Gamma \epsilon_0$}.
\end{equation}
A more detailed account of Bronshtein's proof of~\eqref{goodbronsh}
can be found in Section 8.4 of~\cite{Dudley99book}.  

\begin{IEEEproof}[Proof of Theorem~\ref{difflip}]
The conclusion of the theorem is clearly only meaningful in the case
when $\Gamma_i < \infty$ for all $i = 1, \dots, d$. We therefore
assume this in the rest of this proof.

For every $f \in \C \left(\prod_{i=1}^d [a_i, b_i]; B; \Gamma_1,
  \dots, \Gamma_d \right)$, let us define the function $\hat{f}$ on
$[0, 1]^d$ by
\begin{equation*}
  \hat{f}(t_1, \dots, t_d) := f \left(a_1 + (b_1 - a_1)t_1, \dots, a_d
  + (b_d - a_d) t_d\right),
\end{equation*}
for $t_1,t_2,\ldots,t_d \in [0,1]$. Clearly the function $\hat{f}$
belongs to the class $\C \left([0, 1]^d; B; \Gamma_1(b_1 - a_1),
  \dots, \Gamma_d(b_d - a_d)\right)$ and covering $\hat{f}$ to within 
$\epsilon$ in the $L_{\infty}$-metric is equivalent to covering
$f$. Thus
\begin{eqnarray}\label{prescale}
M \left(\C (\prod_i [a_i, b_i]; B; \Gamma_1, \dots, \Gamma_d), \epsilon; L_{\infty} \right) \;\;\;\;\;\;\;\;\;\;\;\;\;\;\;\;\;\;\;\;\nonumber \\
= M \left(\C ([0, 1]^d; B; \Gamma_1(b_1-a_1), \dots, \Gamma_d(b_d - a_d)), \epsilon;
      L_{\infty} \right).
\end{eqnarray}
We thus take, without loss of generality, $a_i = 0$ and $b_i =
1$ for all $i = 1, \dots, d$. 

From Lemma~\ref{infvf} and the observation that $V_f(B) \in
\K^{d+1}(\sqrt{d + B^2})$ for all $f \in \C([0, 1]^d, B)$, it follows
that 
\begin{eqnarray*}
& M \left(\C ([0, 1]^d; B; \Gamma_1, \dots, \Gamma_d), \epsilon; L_{\infty} \right) \;\;\;\;\;\;\;\;\\
& \leq M \left(\K^{d+1}(\sqrt{d+B^2}), \frac{\epsilon}{2\sqrt{1+\Gamma_1^2 + \dots + \Gamma_d^2}}; \ell_H
\right).
\end{eqnarray*}
Thus from~\eqref{goodbronsh}, we deduce the existence of two positive
constants $c$ and $\epsilon_0$, depending only on $d$, such that
\begin{eqnarray*}
& \log M \left(\C ([0, 1]^d; B; \Gamma_1, \dots, \Gamma_d), \epsilon;
  L_{\infty} \right) \\ &\leq c\left(\frac{\sqrt{(d+B^2)(1+\Gamma_1^2
      + \dots + \Gamma_d^2)}}{\epsilon} \right)^{d/2}, 
\end{eqnarray*}
if $\epsilon \leq \epsilon_0
\sqrt{(d+B^2)(1+\Gamma_1^2+\dots+\Gamma_d^2)}$. By the scaling
inequality~\eqref{prescale}, we obtain
\begin{eqnarray*}
&& \log  M \left(\C (\prod_i[a_i, b_i]; B; \Gamma_1, \dots, \Gamma_d),
  \epsilon; L_{\infty} \right)\;\;\;\;\;\;\; \\ 
&& \;\;\;\leq c \left(\frac{\sqrt{(d+B^2)(1+\sum_i
      \Gamma_i^2(b_i-a_i)^2)}}{\epsilon} \right)^{d/2} 
\end{eqnarray*}
if $\epsilon \leq \epsilon_0 \sqrt{(d+B^2)(1+\sum_i
  \Gamma_i^2(b_i-a_i)^2)}$.
By another scaling argument, it follows that
\begin{eqnarray*}
  && M \left(\C (\prod_i [a_i, b_i]; B; \Gamma_1, \dots, \Gamma_d ),
    \epsilon; L_{\infty} \right) \\
    && = M \left(\C \left(\prod_i[a_i, b_i];
    \frac{B}{\Upsilon}; \frac{\Gamma_1}{\Upsilon}, \dots,
    \frac{\Gamma_d}{\Upsilon} \right), \frac{\epsilon}{\Upsilon};
  L_{\infty} \right)
\end{eqnarray*}
for every $\Upsilon > 0$ and, as a consequence, we get, for every
$\Upsilon > 0$,
\begin{eqnarray*}
&& \log  M \left(\C (\prod_i[a_i, b_i]; B; \Gamma_1, \dots, \Gamma_d),
  \epsilon; L_{\infty} \right)  \\ 
 & \leq & c \left(\frac{\sqrt{(d\Upsilon^2+B^2)(1+\sum_i
        \Gamma_i^2(b_i-a_i)^2/\Upsilon^2)}}{\epsilon} \right)^{d/2}.
\end{eqnarray*}
if $\epsilon \leq \epsilon_0
\sqrt{(d\Upsilon^2+B^2)(1+\sum_i
  \Gamma_i^2(b_i-a_i)^2/\Upsilon^2)}$. Choosing (by differentiation)
\begin{equation*}
  \Upsilon^4 = \frac{B^2\sum_i \Gamma_i^2(b_i-a_i)^2}{d},
\end{equation*}
we deduce finally
\begin{eqnarray*}
    && \log M \left(\C ([a, b]^d; B; \Gamma_1, \dots, \Gamma_d),
      \epsilon; L_{\infty} \right)  \\ & \leq & c 
    \left(\frac{B+\sqrt{d \sum_i \Gamma^2_i(b_i - a_i)^2}}{\epsilon}
    \right)^{d/2}
\end{eqnarray*}
if $\epsilon \leq \epsilon_0 \left(B + \sqrt{d \sum_i
    \Gamma_i^2(b_i-a_i)^2} \right)$. The proof of the theorem will now
be complete by noting that
\begin{equation*}
\sqrt{\sum_i \Gamma_i^2(b_i -
    a_i)^2} \leq \sum_i \Gamma_i(b_i - a_i) \leq  \sqrt{d \sum_i
    \Gamma_i^2(b_i - a_i)^2}.
\end{equation*}
The terms involving $d$ can be absorbed in the constants $c$ and
$\epsilon_0$.
\end{IEEEproof}

One might wonder if a version of Lemma~\ref{lset} can be proved for
the $L_p$-metric instead of the $L_{\infty}$-metric, and without any
Lipschitz constraints. Such an inequality would, in particular, yield
an alternative simpler proof of Theorem~\ref{UpperBound}. It
turns out that one can prove such a bound for the $L_1$-metric but
not for $L_p$ for any $p > 1$. The inequality for $L_1$
is presented next. This inequality could possibly be of independent
interest. The reason why such an inequality can not be proved for
$L_p, p > 1$, is explained in Remark~\ref{nowork}. 
\begin{lemma}\label{lset}
For every pair of functions $f$ and $g$ in $\C([0,1]^d, 1)$, we
have
\begin{equation}\label{lset.eq}
  ||f-g||_1 \leq (1+20d) \ell_H(V_f(1), V_g(1)). 
\end{equation}
\end{lemma}
\begin{IEEEproof}
For $f \in \C([0,1]^d, 1)$ and $x \in (0, 1)^d$, let $m_f(x)$
  denote any subgradient of the convex function $f$ at $x$. Let
  $\ell_H(V_f(1), V_g(1)) = \rho > 0$. Our first step is to observe
  that 
\begin{equation}\label{pwise}
  |f(x) - g(x)| \leq \rho \left(1 + |m_f(x)| +
    |m_g(x)| \right)
\end{equation}
for every $x \in (0, 1)^d$, where $|m_f(x)|$ denotes the Euclidean
norm of the subgradient vector $m_f(x) \in \R^{d}$. To see this, fix
$x \in (0, 1)^d$ with $f(x) \neq g(x)$. We assume, without loss of
generality, that $f(x) < g(x)$. Clearly $(x, f(x)) \in V_f(1)$ and
because $\ell_H(V_f(1), V_g(1)) = \rho$, there exists $(x', y') \in
V_g(1)$ with $|(x, f(x)) - (x', y')| \leq \rho$. Since $f(x) < g(x)$,
the point $(x, f(x))$ lies outside the convex set $V_g(1)$ and we can
thus take $y' = g(x')$. By the definition of the subgradient, we
have $$ g(x') \geq g(x) + \left< m_g(x), x' - x \right>.$$
Therefore,
\begin{align*}
0 \leq  g(x) - f(x) &= g(x) - g(x') + g(x') - f(x) \\
&\leq \left<m_g(x), x - x' \right> + |g(x') - f(x)| \\
&\leq |m_g(x)| |x-x'| + |g(x') - f(x)| \\
&\leq  \sqrt{|m_g(x)|^2 + 1}~ |(x, f(x)) - (x', y')| \\
&\leq \rho \sqrt{|m_g(x)|^2 + 1} \leq \rho (1 + |m_g(x)|).
\end{align*}
Note that the Cauchy-Schwarz inequality has been used twice in the above chain
of inequalities. We have thus shown that $g(x) - f(x) \leq \rho(1 +
|m_g(x)|)$ in the case when $f(x) < g(x)$. One would have a similar
inequality in the case when $f(x) > g(x)$. Combining these two, we
obtain~\eqref{pwise}.

As a consequence of~\eqref{pwise}, we get
\begin{align*}
\;\;\; & ||f - g||_1 \\ 
& =  \int_{[0,1]^d \setminus [\rho,
    1-\rho]^d} |f - g| + \int_{[\rho, 1-\rho]^d} |f - g| \\
&\leq 2\left(1 - (1-2\rho)^d \right) + \rho \left( 1 +
    \int_{[\rho, 1-\rho]^d} |m_f(x)| dx \right. \\
    & \;\;\;\;\;\;\;\;\;\;\;\;\;\;\;\;\;\;\;\;\;\;\;\;\;\;\;\;\;\;\;\;
    \left. + \int_{[\rho,1 - \rho]^d} 
    |m_g(x)| dx   \right) \\
&\leq \rho \left(1+4d +  \int_{[\rho, 1-\rho]^d} \{ |m_f(x)| +
  |m_g(x)|\} dx   \right), 
\end{align*}
where we have used the inequality $(1-2\rho)^d \geq 1-2d\rho$.

To complete the proof of~\eqref{lset.eq}, we show that $\int_{[\rho,
  1-\rho]^d} |m_f(x)|dx \leq 8d$ for every $f \in \C([0,1]^d,
1)$. We write $m_f(x) = (m_f(x)(1), \dots, m_f(x)(d)) \in \R^d$ and
use the definition of the subgradient to note that for every $x \in
[\rho, 1-\rho]^d$ and $1 \leq i \leq d$,
\begin{equation}\label{tmp1}
  f(x + te_i) - f(x) \geq  t \; m_f(x)(i)
\end{equation}
for $t > 0$ sufficiently small, where $e_i$ is the unit vector in the $i$th coordinate direction i.e.,
$e_i(j) := 1$ if $i=j$ and $0$ otherwise. Dividing both sides by $t$ and
letting $t \downarrow 0$, we would get $m_f(x)(i) \leq f'(x;
e_i)$ (we use $f'(x; v)$ to denote the directional derivative of $f$
in the direction $v$; directional derivatives exist as $f$
is convex). Using~\eqref{tmp1} for $t < 0$, we get $m_f(x)(i) \geq
-f'(x; -e_i)$. Combining these two inequalities, we get
\begin{equation*}
  |m_f(x)(i)| \leq |f'(x; e_i)| + |f'(x; -e_i)| \qt{for $i = 1, \dots,
    d$}.
\end{equation*}
As a result,
\begin{eqnarray*}
 &&  \int_{[\rho,1-\rho]^d} |m_f(x)|dx \\
 & \leq & \sum_{i=1}^d
  \int_{[\rho,1-\rho]^d} |m_f(x)(i)| dx \\
  & \leq & \sum_{i=1}^d \left(\int_{[\rho,1-\rho]^d} |f'(x; e_i)| dx
    +   \int_{[\rho,1-\rho]^d} |f'(x; -e_i)| dx \right).
\end{eqnarray*}
We now show that for each $i$, both the integrals $\int_{[\rho,1-\rho]^d}
|f'(x; e_i)|$ and $\int_{[\rho,1-\rho]^d} |f'(x; -e_i)|$ are bounded
from above by 4. Assume, without loss of generality, that $i=1$ and
notice
\begin{eqnarray}\label{edin}
   \int_{[\rho,1-\rho]^d} |f'(x; e_1)| dx \;\;\;\;\;\;\;\;\;\;\;\;\;\;\;\;\;\;\;\;\;\;\;\;\;\;\;\;\;\;\;\;\nonumber \\ 
  = \int_{u \in
    [\rho,1-\rho]^{d-1}} \left( \int_{\rho}^{1-\rho} |f'((x_1,u); e_1)| dx_1
  \right) du.
\end{eqnarray}
We fix $u= (x_2, \dots, x_d) \in [\rho, 1-\rho]^{d-1}$ and focus on the
inner integral. Let $v(z) := f(z, x_2, \dots, x_d)$ for $z \in [0,
1]$. Clearly $v$ is a convex function on $[0, 1]$ and its right
derivative, $v_r'(x_1)$ at the point $z = x_1 \in (0, 1)$ equals
$f'(x; e_1)$ where $x = (x_1, \dots, x_d)$. The inner integral thus
equals $\int_{\rho}^{1-\rho} |v_r'(z)| dz$. Because of the convexity
of $v$, its right derivative $v_r'(z)$ is non-decreasing and satisfies
\begin{equation*}
  v(y_2) - v(y_1) = \int_{y_1}^{y_2} v_r'(z) dz \qt{for $0 < y_1 < y_2
    < 1$}.
\end{equation*}
Consequently,
\begin{eqnarray*}
 &&  \int_{\rho}^{1-\rho} |v_r'(z)| dz \\
  & \leq & \sup_{\rho \leq c \leq
    1-\rho} \left( - \int_{\rho}^c v_r'(z) dz + \int_c^{1-\rho}
    v_r'(z) dz \right) \\
    & = & \sup_{\rho \leq c \leq 1-\rho} \left(v(\rho)
    + v(1-\rho) - 2 v(c) \right).
\end{eqnarray*}
The function $v(\cdot)$ clearly satisfies $|v(z)|
\leq 1$ because $f \in \C([0, 1]^d, 1)$. This implies that
$\int_{\rho}^{1-\rho} |v_r'(z)|dz \leq 4$. The
identity~\eqref{edin} therefore gives
\begin{eqnarray*}
  \int_{[\rho, 1-\rho]^d} |f'(x; e_1)| dx \;\;\;\;\;\;\;\;\;\;\;\;\;\;\;\;\;\;\;\;\;\;\;\;\;\;\;\;\;\\
  = \int_{(x_2, \dots, x_d)
  \in [\rho, 1-\rho]^{d-1}} \left( \int_{\rho}^{1-\rho} |v_r'(z)| dz
\right) dx_2 \dots dx_d \leq 4.
\end{eqnarray*}
Similarly, by working with left derivatives of $v$ as opposed to
right, we can prove that
\begin{equation*}
  \int_{[\rho, 1-\rho]^d} |f'(x; -e_1)| dx \leq 4.
\end{equation*}
Therefore, the integral $\int_{[\rho, 1-\rho]^d}|m_f|$ is at most
$8d$ because it is less than or equal to 
\begin{equation*}
  \sum_{i=1}^d \left(
    \int_{[\rho,1-\rho]^d} |f'(x; e_i)| dx    +
    \int_{[\rho,1-\rho]^d} |f'(x; -e_i)| dx \right). 
\end{equation*}
This completes the proof of Lemma~\ref{lset}.   
\end{IEEEproof}
\begin{remark}\label{nowork}
  Lemma~\ref{lset} is not true if $L_1$ is replaced by $L_p$, for $p >
  1$. Indeed, if $d = 1$ and $f_{\alpha}(x) := \max(0, 1 -
  (x/\alpha))$ for $0 < \alpha \leq 1$ and $g(x) := 0$ for all $x \in
  [0, 1]$, then it can be easily checked that for $1 \leq p < \infty$, 
  \begin{equation*}
    ||f_{\alpha} - g||_p = \frac{\alpha^{1/p}}{(1+p)^{1/p}} \text{ and
    } \ell_H(V_{f_{\alpha}}(1), V_g(1)) = \frac{\alpha}{\sqrt{1+\alpha^2}}. 
  \end{equation*}
As $\alpha$ can be arbitrarily close to zero, this clearly rules out
any inequality of the form~\eqref{lset.eq} with the $L_1$-metric
replaced by $L_p$, for $1 < p \leq \infty$.  
\end{remark}
\begin{remark}
Lemma~\ref{lset} and Bronshtein's result~\eqref{goodbronsh} can be
used to give an alternative proof of Theorem~\ref{UpperBound} for the
special case $p = 1$. Indeed, the scaling identity~\eqref{uglyscale}
lets us take $a = 0$, $b = 1$ and $B = 1$. Inequality~\eqref{lset.eq}
implies that the covering number $M \left(\C([0, 1]^d, 1), \epsilon;
  L_1 \right)$ is less than or equal to 
\begin{equation*}
M \left(\K^{d+1}(\sqrt{d+1}), \frac{\epsilon}{2(1+20d)}; \ell_H
\right).
\end{equation*}
Thus from~\eqref{goodbronsh}, we deduce the existence of two positive
constants $c$ and $\epsilon_0$, depending only on $d$, such that
\begin{equation*}
\log M \left(\C ([0, 1]^d, 1), \epsilon; L_1 \right)  \leq c
\epsilon^{-d/2}
\end{equation*}
whenever $\epsilon \leq \epsilon_0$. Note that, by
Remark~\ref{nowork}, this method of proof does not work in the case of
$L_p$, for $1 < p < \infty$. 
\end{remark}





\appendices
\ifCLASSOPTIONcaptionsoff
  \newpage
\fi



%

\bibliographystyle{IEEEtran}
\bibliography{AG}

\def\noopsort#1{}
\begin{thebibliography}{10}
\providecommand{\url}[1]{#1}
\csname url@samestyle\endcsname
\providecommand{\newblock}{\relax}
\providecommand{\bibinfo}[2]{#2}
\providecommand{\BIBentrySTDinterwordspacing}{\spaceskip=0pt\relax}
\providecommand{\BIBentryALTinterwordstretchfactor}{4}
\providecommand{\BIBentryALTinterwordspacing}{\spaceskip=\fontdimen2\font plus
\BIBentryALTinterwordstretchfactor\fontdimen3\font minus
  \fontdimen4\font\relax}
\providecommand{\BIBforeignlanguage}[2]{{%
\expandafter\ifx\csname l@#1\endcsname\relax
\typeout{** WARNING: IEEEtran.bst: No hyphenation pattern has been}%
\typeout{** loaded for the language `#1'. Using the pattern for}%
\typeout{** the default language instead.}%
\else
\language=\csname l@#1\endcsname
\fi
#2}}
\providecommand{\BIBdecl}{\relax}
\BIBdecl

\bibitem{KolmogorovTihomirov}
A.~N. Kolmogorov and V.~M. Tihomirov, ``$\epsilon$-entropy and
  $\epsilon$-capacity of sets in function spaces,'' \emph{Amer. Math. Soc.
  Transl. (2)}, vol.~17, pp. 277--364, 1961.

\bibitem{Dryanov}
D.~Dryanov, ``Kolmogorov entropy for classes of convex functions,''
  \emph{Constructive Approximation}, vol.~30, pp. 137--153, 2009.

\bibitem{Bronshtein76}
E.~M. Bronshtein, ``$\epsilon$-entropy of convex sets and functions,''
  \emph{Siberian Mathematical Journal}, vol.~17, pp. 393--398, 1976.

\bibitem{Dudley99book}
R.~M. Dudley, \emph{Uniform Central Limit Theorems}.\hskip 1em plus 0.5em minus
  0.4em\relax Cambridge University Press, 1999.

\bibitem{Birge83}
L.~Birg{\'e}, ``Approximation dans les espaces metriques et theorie de
  l'estimation,'' \emph{Zeitschrift f{\"u}r Wahrscheinlichkeitstheorie und
  Verwandte Gebiete}, vol.~65, pp. 181--237, 1983.

\bibitem{LeCam:73AnnStat}
L.~Le~Cam, ``Convergence of estimates under dimensionality restrictions,''
  \emph{Annals of Statistics}, vol.~1, pp. 38--53, 1973.

\bibitem{YangBarron}
Y.~Yang and A.~Barron, ``Information-theoretic determination of minimax rates
  of convergence,'' \emph{Annals of Statistics}, vol.~27, pp. 1564--1599, 1999.

\bibitem{GuntuFdiv}
A.~Guntuboyina, ``Lower bounds for the minimax risk using $f$ divergences, and
  applications,'' \emph{IEEE Transactions on Information Theory}, vol.~57, pp.
  2386--2399, 2011.

\bibitem{VandegeerBook}
S.~Van~de Geer, \emph{Applications of Empirical Process Theory}.\hskip 1em plus
  0.5em minus 0.4em\relax Cambridge University Press, 2000.

\bibitem{BM93}
L.~Birg\'e and P.~Massart, ``Rates of convergence for minimum contrast
  estimators,'' \emph{Probability Theory and Related Fields}, vol.~97, pp.
  113--150, 1993.

\bibitem{SS11}
E.~Seijo and B.~Sen, ``Nonparametric least squares estimation of a multivariate
  convex regression function,'' \emph{Annals of Statistics}, vol.~39, pp.
  1633--1657, 2011.

\bibitem{HD11}
L.~A. Hannah and D.~Dunson, ``Bayesian nonparametric multivariate convex
  regression,'' 2011, submitted.

\bibitem{SW10}
A.~Seregin and J.~A. Wellner, ``Nonparametric estimation of multivariate
  convex-transformed densities,'' \emph{Annals of Statistics}, vol.~38, pp.
  3751--3781, 2010.

\bibitem{CSS10}
M.~L. Cule, R.~J. Samworth, and M.~I. Stewart, ``Maximum likelihood estimation
  of a multi-dimensional log-concave density (with discussion),'' \emph{Journal
  of the Royal Statistical Society, Series B}, vol.~72, pp. 545--600, 2010.

\bibitem{DSS11}
L.~D\"umbgen, R.~J. Samworth, and D.~Schuhmacher, ``Approximation by
  log-concave distributions with applications to regression,'' \emph{Annals of
  Statistics}, vol.~39, pp. 702--730, 2011.

\bibitem{G11}
A.~Guntuboyina, ``Optimal rates of convergence for the estimation of
  reconstruction of convex bodies from noisy support function measurements.''
  \emph{Annals of Statistics}, 2011, to appear.

\bibitem{GKM06}
R.~J. Gardner, M.~Kiderlen, and P.~Milanfar, ``Convergence of algorithms for
  reconstructing convex bodies and directional measures,'' \emph{Annals of
  Statistics}, vol.~34, pp. 1331--1374, 2006.

\bibitem{Massart03Flour}
P.~Massart, \emph{Concentration inequalities and model selection. Lecture notes
  in Mathematics}.\hskip 1em plus 0.5em minus 0.4em\relax Berlin: Springer,
  2007, vol. 1896.

\end{thebibliography}
\def\noopsort#1{}

%








\end{document}